\documentclass[aps,pra,showpacs,twocolumn]{revtex4}
\usepackage[pdftex]{graphicx}
\usepackage{amssymb}
\usepackage{amsmath}
\usepackage{bm}

\begin{document}

\title{Scaling behavior of density fluctuations in an expanding quasi-2D degenerate Bose gas}

\author{Sang Won Seo, Jae-yoon Choi, and Yong-il Shin}\email{yishin@snu.ac.kr}

\affiliation{Center for Subwavelength Optics and Department of Physics and Astronomy, Seoul National University, Seoul 151-747, Korea}

\begin{abstract}
We measure the power spectrum of density fluctuations emerged in a freely expanding quasi-two-dimensional (2D) degenerate Bose gas and investigate the scaling behavior of the spectrum for the expansion time. The power spectrum shows an oscillatory shape for long expansion times, where the spectral peak positions are observed to be shifted to lower spatial frequencies than the theoretical prediction for a non-interacting expansion case. We find the spectral peak positions in good agreement with the recent numerical simulation presented by Mazets [Phys.~Rev.~A \textbf{86}, 055603 (2012)], where the atom-atom interactions are taken into account. We present a mean-field description of the interaction effect in the expansion dynamics and quantitatively account for the observed spectral peak shifts. The spectral shift is intrinsic to the free expansion of a quasi-2D Bose gas due to finite axial confinement. Finally, we investigate the defocussing effect in the power spectrum measurement.
\end{abstract}


\maketitle

\section{Introduction}

In a two-dimensional (2D) Bose gas, Bose-Einstein condensation is prohibited at finite temperature due to large thermal fluctuations~\cite{Mermin,Hohenberg}. However, the system can show superfluidity below a certain critical temperature, where the first-order spatial correlation function decays algebraically for large distances as $g_1(r)\sim r^{-\eta}$. Since no specific correlation length is involved in such a slow decay, the 2D superfluid state exhibits quasi-long-range order. At low temperature, the decay exponent is given as $\eta=1/(n_s\lambda^2)$, related with the superfluid density $n_s$ and the thermal de Broglie wavelength $\lambda$. The Berezinskii-Kosterlitz-Thouless (BKT) theory predicts that $\eta$ is increased up to a universal value of 1/4 as the temperature approaches to the critical point~\cite{Berezinskii,KT}. The superfluid-to-normal phase transition is associated with proliferation of free vortices, which transforms the decay behavior of the phase coherence from algebraic to exponential.

Cold atomic systems provide a versatile platform to study the microscopic mechanism for 2D superfluidity and the BKT physics. It has been experimentally demonstrated that a BKT-type phase transition occurs in a finite-size quasi-2D Bose gas trapped in a harmonic potential~\cite{Hadzibabic,Clade,Tung,Hung,Plisson,Yefsah}. Recently, superfluid behavior was demonstrated by measuring a critical velocity~\cite{Desbuquois} and thermally activated vortex pairs were directly observed~\cite{ThermalVP}. The power-law decay behavior of the 2D superfluid state has been investigated using interferometric techniques~\cite{Hadzibabic,Clade}. However, quantitative tests of the BKT predictions on the decay exponent in the superfluid state have not been reported yet. In particular, experimental verification of the relation between the superfluid density and the decay exponent is highly desirable.

Another interesting method to probe the phase correlations in atomic samples is measuring density correlations in freely expanding samples~\cite{Altman}. For low-dimensional quasicondensates, phase fluctuations in the initial samples develop into density modulations during expansion as a result of near-field diffraction of matter wave. This method has been successfully employed for the study of enlongated condensates~\cite{Dettmer} and one-dimensional quasicondensates~\cite{Imambekov,Manz}. Particularly for a homogeneous 2D degenerate Bose gas below the BKT transition, Imambekov \textit{et al.}~\cite{Imambekov} showed that the power spectrum of the density fluctuations has a self-similar, oscillatory shape for sufficiently long expansion times. This self-similarity is a clear manifestation of the power-law decay of the phase coherence. Furthermore, the self-similar spectral shape depends only on the decay exponent $\eta$, suggesting a new route for quantitative determination of $\eta$~\cite{Imambekov,Mathey}.

In this paper, we study the power spectrum of density fluctuations in an expanding quasi-2D degenerate Bose gas. In our previous experiment~\cite{PhaseFluc}, we confirmed the thermal nature of the phase fluctuations by investigating the temperature dependence of the spectral strength, and demonstrated the usage of the power spectrum for studying nonequilibrium dynamics. However, the scaling behavior of the power spectrum was much different from the theoretical prediction in Ref.~\cite{Imambekov}. In subsequent works~\cite{JKPS_Seo}, we found that the imaging setup in the experiment was subject to defocussing due to the gravitational free fall of the sample, which can severely affect the measurement of density fluctuations~\cite{LangenComment}.

Here we present the power spectrum of density fluctuations measured by optimally focused imaging and investigate the scaling behavior of the spectrum for the expansion time. The power spectrum exhibits an oscillatory shape for long expansion times as expected with the power-law decay of the phase coherence. We find that the spectral peak positions are shifted from those predicted in Ref.~\cite{Imambekov} but in good agreement with the recent numerical simulations including the atom-atom interactions in the expansion dynamics~\cite{Igor}. We present a mean-field description of the interaction effect in the expansion dynamics and quantitatively account for the observed spectral shifts. We also show that the shift is intrinsic to the free expansion of a quasi-2D Bose gas with finite $\mu_0/\hbar\omega_z$, where $\mu_0$ is the chemical potential and $\hbar\omega_z$ is the axial confinement energy. Determination of the decay exponent from the spectral shape is precluded in our experiment, calling for further theoretical investigation on the expansion dynamics.

This paper is organized as follows: In Sec.~II we describe the experiment procedure to measure the power spectrum and in particular, how we optimize the imaging focus to a free falling sample. In Sec.~III we present experiment results and discuss on the interaction effect in the scaling behavior of the power spectrum. In Sec.~IV we investigate the defocussing effect in the power spectrum measurement. We summarize our results in Sec.~V.

\section{Experiment}

Our experimental setup and sample preparation procedure were described in Ref.~\cite{SNUBEC,PhaseFluc}. We generate thermal $^{23}$Na atoms in the $|F=1,m_F=-1\rangle$ state in an optically plugged magnetic quadrupole trap, and transfer them into a single pancake-shape optical dipole trap. A degenerate Bose gas is generated by applying evaporation cooling with slowly ramping down the optical trap depth. The trapping frequencies of the optical trap are $(\omega_x, \omega_y, \omega_z) = 2\pi \times$(3.0, 3.9, 370)~Hz, where the $z$ direction is along the gravity. The total atom number and temperature of the sample are $N_t\approx 1.2\times 10^6$ and $T\approx 50$nK, respectively. The coherent part fraction is about 0.5, which is determined from a bimodal fit to the density distribution after expansion. The \textit{in situ} radius of the coherent part is measured to be $R_{y}\approx 105~\mu$m~\cite{ThermalVP}, giving an estimate of the chemical potential $\mu_0= m\omega_{y}^2 R_{y}^2 /2 \approx h\times 190$~Hz, which is less than the confining energy $\hbar \omega_z$. We note that $k_B T/\hbar\omega_z\approx 3$ and thermal populations in the axial direction is not negligible in our sample. The dimensionless interaction strength $g=a\sqrt{8\pi m \omega_z / \hbar} \simeq 0.013$, where $a$ is the 3D scattering length.

Free expansion of the sample is initiated by turning off the optical potential. After an expansion time $t_e$, we measure the column density distribution $n(\vec{r};t_e)$ of the sample by taking an absorption image along the $z$ direction. In our setup, the sample falls along the imaging axis due to the gravity when it is released from the trap. Because the depth of field is about 50~$\mu$m for our imaging resolution of $\approx 8~\mu$m, the defocussing due to the free fall becomes nonnegligible for $t_e>3$~ms and the recorded image would be the near-field diffraction pattern of a probe beam after the sample.

\begin{figure}
\includegraphics[width=8.3cm]{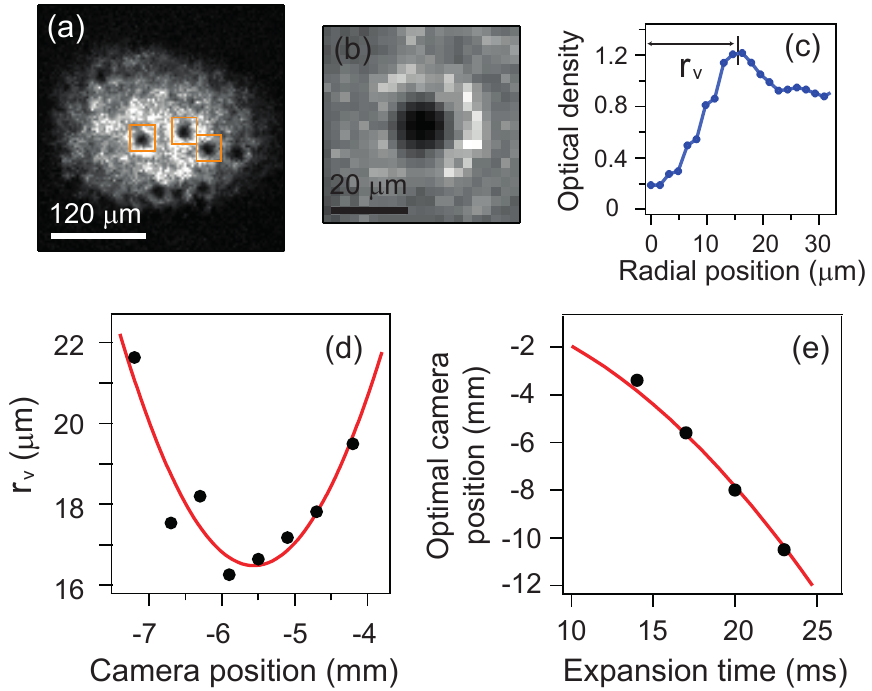}
\caption{
(Color online) (a) Optical density image of a condensate with quantized vortices after 17-ms of time-of-flight. (b) Average image of the vortex core regions which are indicated by red boxes in (a). (c) Radial profile of the density-depleted vortex core is obtained by radially averaging the image in (b). $r_v$ is the radial position of the first peak in the profile. (d) $r_v$ measured for various axial positions of the camera in the imaging setup. The optimal camera position $z_c^*$ is determined for the minimum $r_v$ from a parabola fit (red line) to the data. (e) $z_c^*$ as a function of the expansion time $t_e$. The solid line is a fit line of $z_c^*=4.02(7)\times(g t_e^2/2)$, where $g$ is the gravitational acceleration.}
\label{fig1}
\end{figure}

We compensate the free fall distance by adjusting the axial position $z_c$ of the camera in the imaging setup. In order to find the optimal position $z_c^*$ of the camera, we take advantage of condensates containing quantum vortices, which have a spatially small structure of density-depleted vortex cores as a reference to imaging focus. We prepared almost pure condensates and generated vortices by mechanically perturbing the condensates with the optical plug beam~\cite{PhaseFluc,JKPS_Seo}. For a given expansion time $t_e$, we took images for various camera positions and determined $z_c^*(t_e)$ for minimizing the measured radius of the vortex core (Fig.~1). This imaging optimization works only for $t_e>10$~ms when the vortex core visibility is reasonably high. A parabolic fit to $z_c^*(t_e)$ gives $z_c^*=4.02(7)\times (gt_e^2/2)$ which is well explained with the free fall distance $d=gt_e^2/2$ and the magnification ratio $M=2.0$ of our imaging setup ($g$ is the gravitational acceleration). 

In this work, we restrict the expansion time $t_e\leq 22$~ms for which the thickness of the expanding cloud is not significant larger than the depth of field of our imaging. For longer expansion times, one may employ a spatial pumping technique to image only a cross section of the sample~\cite{VLattice}. In the far-field diffraction limit, $t_e\gg m R_{x,y}^2/h\sim1$~s, the density distribution reveals the momentum distribution of the sample~\cite{Plisson,MatheyPRA}.

\begin{figure}
\includegraphics[width=8.3cm]{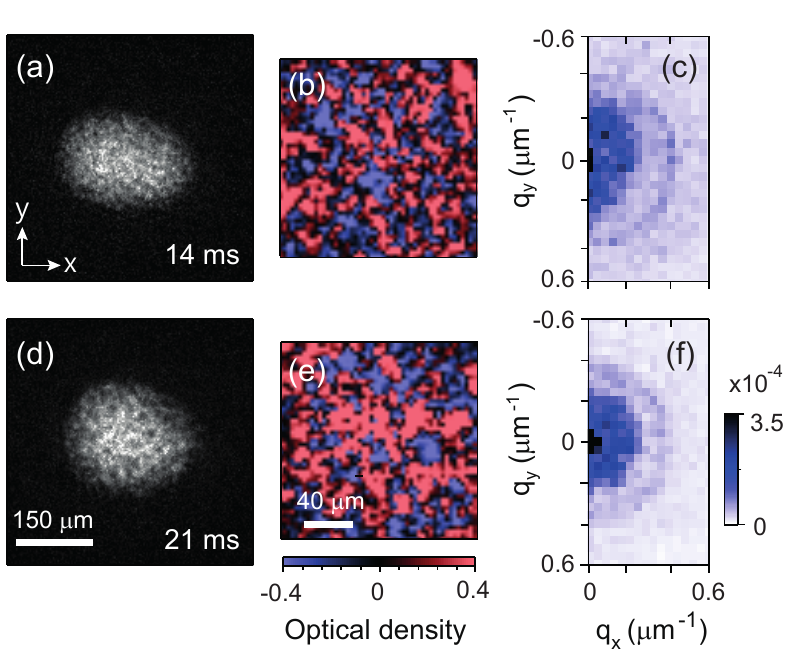}
\caption{
(Color online) Images of quasi-2D condensates after expanding for (a) $t_e=14$~ms and (d) 21~ms. Density fluctuations emerge in the course of expansion. (b) and (e) are the density fluctuation distributions, $\delta n(\vec{r})=n(\vec{r})-\bar{n}(\vec{r})$, in the center region of the sample for (a) and (d), respectively. Here, $\bar{n}(\vec{r})$ is the average density distribution obtained from 30 realizations of the same experiment. Power spectra $P_e(\vec{q};t_e)$ of density fluctuations at (c) $t_e=14$~ms and (f) 21~ms.}
\label{fig2}
\end{figure}

\begin{figure}
\includegraphics[width=7.5cm]{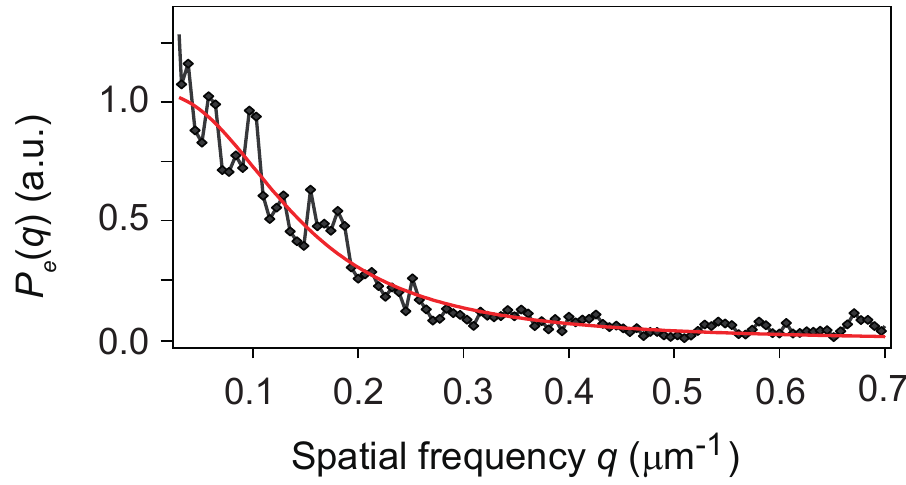}
\caption{
(Color online) Power spectrum of a trapped thermal gas at high temperature. The solid line is a model fit to the spectrum, giving the relative modulation transfer function $M^2(q)$ of our imaging system (see text for detail).
}
\label{fig3}
\end{figure}

Density fluctuations are observed to develop in the expanding sample (Fig.~2). Fig.~2(b) and (e) show examples of density fluctuations distributions, $\delta n(\vec{r})=n(\vec{r})-\bar{n}(\vec{r})$, where $\bar{n}(\vec{r})$ is the average density distribution of many realizations of the same experiment. The spatial pattern of density fluctuations appears random in each realization and there are no noticeable features in the averaged image. This excludes the possibility of uncontrolled modulations in the trapping potential~\cite{Aspect}.

We study the spatial density-density correlations with the normalized power spectrum $P_e(\vec{q})$ of the density fluctuation distribution $\delta n(\vec{r})$, which is defined as 
\begin{equation}
P_{e}(\vec{q};t_e)= \langle \frac{1}{N^2}\big|\int_\mathcal{R} d^2\vec{r} ~e^{i\vec{q}\cdot\vec{r}}\delta n(\vec{r};t_e)\big|^{2}\rangle,
\end{equation}
where $\mathcal{R}$ is the analysis region set to be a 160$\times$160~$\mu$m$^2$ (50$\times$50 pixel$^2$) rectangular region in the center of the sample and $N=\int_\mathcal{R} d^2\vec{r}~\bar{n}(\vec{r})$ is the total atom number. Over the analysis region, the local chemical potential varies about 40\%. We obtain $P_e(\vec{q};t_e)$ from 30 image data for each expansion time $t_e$ and acquire an 1D spectrum $P_e(q;t_e)$ by radially averaging the 2D spectrum. Because the spatial size of one pixel in the image is smaller than the imaging resolution, the spectral signal at high spatial frequency $q>1~\mu$m$^{-1}$ is purely contributed from photon shot noises. We subtract the constant value at high $q$ from the power spectrum. 

The modulation transfer function $\mathcal{M}^2(\vec{q})$ of our imaging system is calibrated by measuring $P_e(\vec{q})$ for trapped thermal gases at high temperature~\cite{ChinNJP}. When the thermal wavelength $\lambda$ is shorter than the imaging resolution, density fluctuations in the thermal gas appear uncorrelated in the image. Therefore, $P_e(\vec{q})$ would linearly reveal $\mathcal{M}^2(\vec{q})$. From a model fit to $P_e (q)$ for the thermal gases, we estimate $\mathcal{M}^2(q)=(1+102\times q^{2.33})^{-1}$ for our imaging (Fig.~3)~\cite{footnote1}. Finally, we determine the power spectrum of density fluctuations as $P(q;t_e)=P_e(q;t_e)/\mathcal{M}^2(q)$.

\section{Result}

\subsection{Spectral peak positions}

\begin{figure*}
\includegraphics[width=11cm]{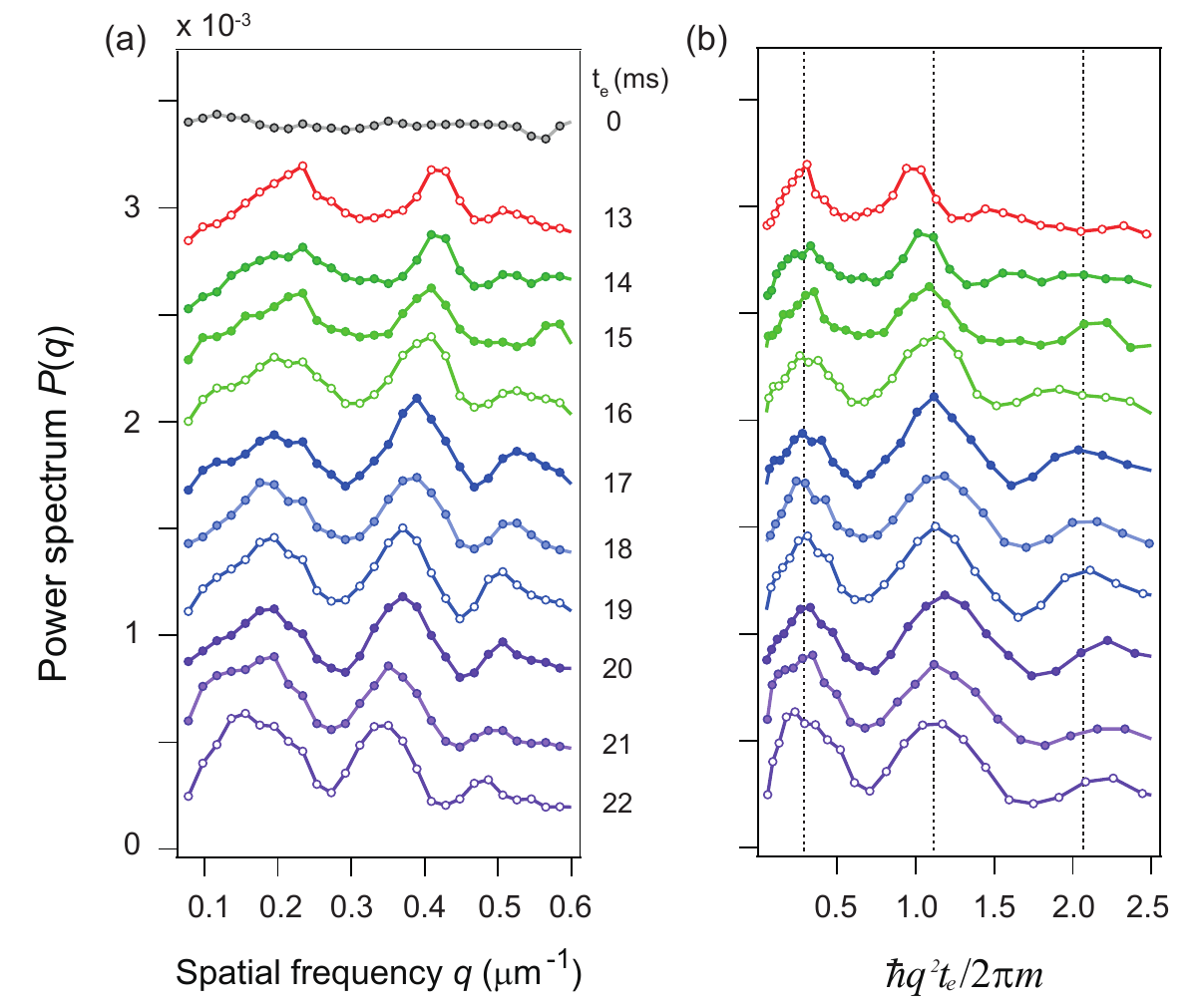}
\caption{
(Color online) (a) Temporal evolution of the power spectrum of density fluctuations. Each spectrum is displayed with an offset for clarity. (b) The same spectra are replotted as functions of a dimensionless parameter $\hbar q_n^2 t_e/ 2 \pi m$. The dotted vertical lines indicate the theoretical prediction for the spectral peak positions from Ref.~\cite{Igor}. 
}
\label{fig4}
\end{figure*}

For the case of the non-interacting expansion of a homogeneous quasicondensate, the theory in Ref.~\cite{Imambekov} predicts development of an oscillatory power spectrum, where the $n$th peak position, $q_n$ ($n=1,2,\cdots$), closely satisfies the relation
\begin{equation}
\hbar q_n^2 t_e/ 2 \pi m \simeq n-1/2.
\end{equation}
As a heuristic example to catch the physical meaning of this relation, let's consider the time evolution of a wave function $\psi(\vec{r})=1+i|\delta\psi|\cos (\vec{q}\cdot\vec{r})$ with $|\delta\psi|\ll1$. This is a superposition of three momentum states of 0, $+\hbar\vec{q}$, and $-\hbar\vec{q}$, and has small density modulations of order of $|\delta\psi|^2$ at $t=0$. After free propagation for time $t$, the wave function evolves into $\psi(\vec{r},t)=1+i|\delta\psi|\cos (\vec{q}\cdot\vec{r})e^{-i\phi (q,t)}$, where $\phi(q,t)=\hbar q^2 t/2m$. The density modulations develop as $\delta n(t)=2|\delta \psi| \cos(\vec{q}\cdot\vec{r}) \sin \phi + O(|\delta \psi|^2)$. Here, we see that the relation in Eq.~(2) corresponds to $\phi(q,t)=(n-1/2)\pi$, which is the condition for maximum density modulations in the above example. The oscillation of the power spectrum can be also described as a Talbot effect in near-field diffraction of matter wave~\cite{PhaseFluc,Imambekov,Talbot}.

Figure 3 displays the power spectra measured for various expansion times. It is clearly observed that an oscillatory spectrum emerges in the course of expansion. In order to compare the scaling behavior of the spectral peak positions to Eq.~(2), we replot the spectra in Fig.~3(b) as functions of the dimensionless parameter $\hbar q^2 t_e/2\pi m$. Apparently, the peak positions in our spectrum does not satisfy the relation in Eq.~(2), but are shifted to lower spatial frequencies. Recently, Mazet~\cite{Igor} performed numerical simulations of the 2D expansion of a degenerate Bose gas taking into account the realistic condition of our experiment and found that the atom-atom interaction effect is not negligible in the early stage of the expansion. From the simulation results, he suggested a relation between $q_n$ and $t_e$ as $\hbar q_n^2 t_e/ 2 \pi m \simeq 0.9 (n-0.7)$. Our experiment results are indeed in quite good agreement with this prediction, demonstrating the significance of the interaction effect in the expansion dynamics of a quasi-2D Bose gas. 

As the trapping potential is released, the quasicondensate expands fast along the axial direction, but with a finite speed. Because the interaction energy is not immediately quenched off but gradually decreases, the aforementioned accumulated phase $\phi_q$ for a momentum state would be larger than that in the non-interacting expansion case. Then, for a given expansion time the interference condition for maximum density modulations would be satisfied with a smaller momentum state. This explains the observed shift direction of the spectral peak positions.

In a mean-field description, we can make quantitative estimation on the spectral shift. First, assuming that the axial expansion of the sample is not affected by its slow transverse expansion, the axial width of the sample increases proportionally to $\sqrt{1+(\omega_z t_e)^2}$. For reducing three-dimensional local density, the interaction energy decreases as $\mu(t)=\mu_0/\sqrt{1+(\omega_z t_e)^2}$. The eigenstate of momentum $\hbar q$ is a phonon excitation with an energy of $\epsilon_q=\sqrt{\epsilon_q^0(\epsilon_q^0+2\mu)}$, where $\epsilon_q^0=\hbar^2 q^2/2m$. Under the assumption that the phonon excitation adiabatically transforms into the free particle state of the same momentum in the expansion dynamics, we estimate the accumulated phase $\phi$ for the momentum state as $\phi(q,t_e)=\int^{t_e}_0 \epsilon_q(t)/\hbar~dt$. From the condition $\phi (q_n,t_e)=(n-1/2)\pi$, we have a modified relation as
\begin{equation}
\frac{\hbar q_n^2}{2 \pi m} \int^{t_e}_0 \sqrt{ 1+\frac{2\mu(t)}{ \epsilon_{q_n}^0} }~dt = n-\frac{1}{2},
\end{equation}
including the time-varying interaction effect. This relation becomes identical to Eq.~(2) when $\mu(t)=0$.

To check the validity of this mean-field model, we calculate $\phi(q_n,t_e)$ for the peak positions $q_n$ in the measured spectra (Fig.~5). The peak positions are determined from a multiple Gaussian fit to the power spectra $P(q)$ and the chemical potential is set to be the average local chemical potential $\bar{\mu}_0=h\times 128$~Hz. Remarkably, $\phi_{q_{1,2,3}}$ are found to be almost constant over the whole range of the expansion time in our experiment and furthermore, their average values are $\bar{\phi}_{q_{1,2,3}}/\pi= 0.54, 1.48$, and 2.57, respectively, which are very close to the values in Eq.~(3). This observation shows that the mean-field model provides a valid picture of the interaction effect on the spectral peak shift in the power spectrum.

It might be expected that the interaction effect would be suppressed with increasing the axial trapping frequency. To see the relative significance of the interaction effect in the spectral peak shift, we express the phase $\phi(q,t_e)$ in terms of dimensionless quantities, $\tilde{q}=\sqrt{\hbar/2m\omega_z}q$ and $\tilde{t}_e=\omega_z t_e$, giving 
\begin{equation}
\phi(q,t_e)=\tilde{q}^2 \int^{\tilde{t}_e}_0 \sqrt{ 1+\frac{2 (\mu_0/\hbar \omega_z)}{\tilde{q}^2\sqrt{1+\tilde{t}^2}} }~d\tilde{t}.
\end{equation}
For long expansion time $\tilde{t}_e\gg1$, the relative shift of the spectral peak positions is determined only by $\mu_0/\hbar\omega_z$, which is a parameter representing the `2D-ness' of the quasi-2D system. For a fixed areal density, $\mu_0/\hbar\omega_z \propto \omega_z^{-1/2}$. The spectral shift due to the interaction effect is intrinsic to the free expansion of a quasi-2D Bose gas due to finite axial trapping frequency. 

\begin{figure}
\includegraphics[width=6.6cm]{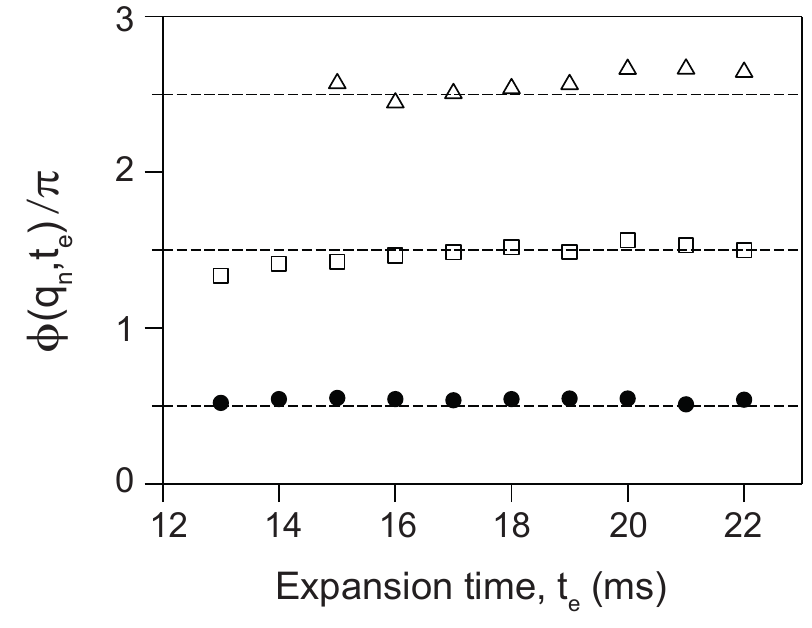}
\caption{
(Color online) Accumulated phase $\phi(q_n,t_e)$ calculated for the spatial frequencies $q_1$ (solid circle), $q_2$ (open square), and $q_3$ (open triangle) of the first, second, and third spectral peaks, respectively, in the power spectrum data.
}
\label{fig5}
\end{figure}

\subsection{Spectral shape}

In the non-interacting expansion case, it is suggested that the power-law decay exponent can be determined from the spectral shape and/or the growth rate of the spectral strength~\cite{Imambekov,Mathey}. Although we observed that the interaction effect significantly affect the power spectrum during the expansion, it is still of high interest to look into  whether there is any reminiscence of the self-similarity and characterize the modified spectral shape.

Before discussing the spectral shape, we have to mention about background noises in our measurement. We observe that the measured spectrum $P(q)$ shows oscillatory behavior with a large background offset. The magnitude of the offset is almost same for all expansion times including $t_e=0$~ms [Fig.~4(a)]. One of the possible sources for the offset is the thermal component coexisting with the quasicondensate. With $k_B T/\hbar\omega_z \sim 3$, there are non-negligible thermal populations in the axially excited states. Because at $T=50$~nK $\lambda \approx1.6~\mu$m is still shorter than the imaging resolution, the contribution of the thermal component appears as a constant offset to the power spectrum. However, the observed offset value is too large to be accounted for with the thermal populations. We found that this anomalous offset becomes more noticeable for higher optical depth~\cite{footnote1} and its magnitude is independent of the sample temperature, but could not identify the origin of the noises yet. In the following, we simply focus on the oscillatory part of the power spectrum. 

We characterize the evolution of the spectra shape with a parameter $S=[P(q_2)-P_m]/[P(q_1)-P_m]$ which is the ratio of the spectral strengths of the first and the second peaks with respect to the minimum value $P_m$ of $P(q)$ between the two peaks (the inset of Fig.~6). $S(t_e)$ slightly increases from 1 to 2 for $t_e<18$~ms and then decreases below 1. In Fig.~6, we display the overlap of the scaled power spectra $[P(q)-P_m]/[P(q_1)-P_m]$, showing that the width of the second spectral peak increases and becomes saturated after $t\geq 18$~ms. The variations of $S$ and the spectral width indicate that the power spectrum is not self-similar in our experiment. The atom-atom interactions during the expansion can result in modification of the spectrum shape as observed in the spectral peak shift. Note that $S$ is expected to be less than 1 in the non-interacting expansion case~\cite{Imambekov, Mathey}. Our results call for further theoretical investigation on the interaction-induced modification of the spectral shape. 

\begin{figure}
\includegraphics[width=6.7cm]{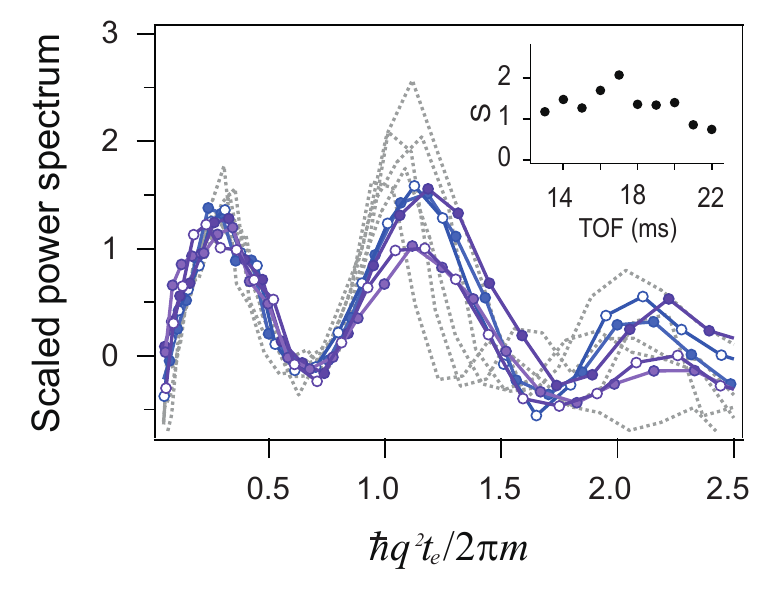}
\caption{
(Color online) Scaled power spectrum $[P(q)-P_m]/[P(q_1)-P_m]$ as a function of the parameter $\hbar q_n^2 t_e/ 2 \pi m$. $P_m$ is the minimum value of $P(q)$ for $q_1<q<q_2$. The symbols are the same as in Fig.~4 and the gray dotted lines are the spectra for $t_e<18$~ms. The inset shows $S=[P(q_1)-P_m]/[P(q_2)-P_m]$.  
}
\label{fig6}
\end{figure}

\section{Defocussing effect}

\begin{figure}
\includegraphics[width=7.0cm]{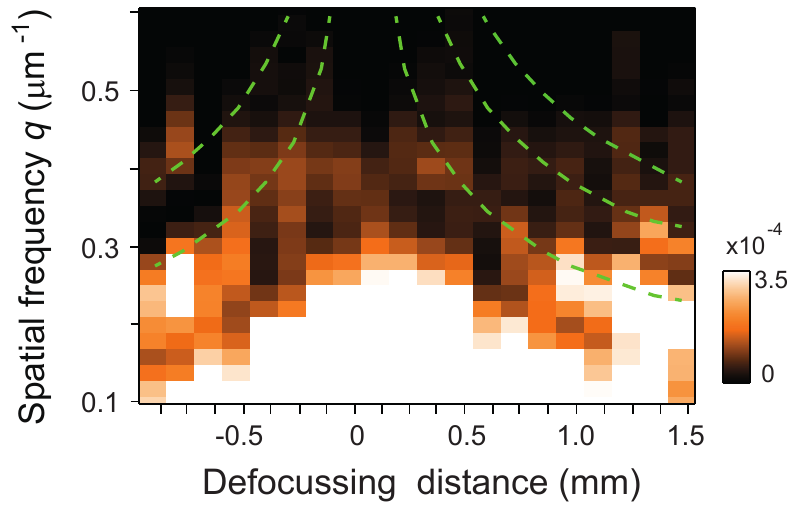}
\caption{
(Color online) Defocussing effect in the power spectrum measurement. Power spectra were measured for various camera positions $z_c$ near the optimal position $z_c$. The defocussing distance $d=(z_c-z_c^*)/M^2$. The dashed lines indicate the spectral peak positions calculated from Eq.~(5).}
\label{fig7}
\end{figure}

In this section, we investigate the defocus effect on the power spectrum measurement. When the camera position is displaced from the image plane of the sample, the recorded image on the camera is the intensity distribution of a probe beam after propagating by the defocussing distance after the sample. Here, the fluctuated density distribution of the sample acts as an intensity mask for the probe beam and the beam propagation would result in distortion of the power spectrum.

Figure 7 shows the power spectra obtained at $t_e=17$~ms for various camera positions around its optimal position $z_c^*$. As the defocussing distance $d=|z_c-z_c^*|/M^2$ is increased, the spectrum becomes more oscillatory with the peak positions $q_n$ monotonically decreasing. We confirmed that the power spectrum is recovered to that reported in our previous work with $d\approx g t_e^2/2$~\cite{PhaseFluc}. We studied the defocussing effect also with numerical simulations of the probe beam propagation. First, assuming a thin sample, we constructed an artificial intensity distribution of the probe beam right after the sample as $I(\vec{r},d=0)=\exp(-\sigma \mathcal{F}^{-1}[N\sqrt{P(|\vec{q}|)}e^{i\theta(\vec{q})}])$, where $\sigma$ is the cross section of the atom, $\mathcal{F}^{-1}$ is the inverse Fourier transform, and $\theta(\vec{q})$ is a randomly generated phase distribution. The intensity distribution $I(\vec{r},d)$ of the probe beam after propagating by a distance $d$ was calculated by using the angular spectrum method~\cite{Goodman}, and the resultant power spectrum with the defocussing effect was obtained from $I(\vec{r},d)$. We confirmed the same behavior of the spectral peak positions in the numerical simulations.

In a recent study on imaging focus optimization~\cite{Putra}, it is shown that the defocussing effect in the power spectrum of a thin sample can be approximately expressed as 
\begin{equation}\label{eq6}
\begin{split}
P_e(q;d)=P_e(q;d=0)\times \cos^2{(\frac{q^2d}{2k_0})},
\end{split}
\end{equation}
where $k_0$ is the wavenumber of the probe beam. We find that our experimental data are satisfied with this relation (Fig.~7). The last factor, $\cos^2{(q^2d/2k_0)}$ becomes zero when $q^2d/2k_0=(n-1/2)\pi$, which has a close correspondence to Eq.(2). This analogy stems from the fact that the structure of the Helmholtz equation of the light propagation is identical to that for the two-dimensional expansion dynamics of non-interacting matter wave.

\section{Summary}

We have investigated the scaling behavior of density fluctuations in an expanding quasi-2D Bose gas by measuring their power spectrum. The spectral peak positions in the power spectrum were observed to be shifted to lower spatial frequencies from the theoretical prediction for the ideal case of non-interacting expansion. We presented a mean-field description of the interaction effect during the expansion and quantitatively accounted for the spectral peak shift. In our experiment, self-similarity of the spectrum shape for long expansion times was not observed and determination of the power-law decay exponent of the phase coherence was precluded. 

Most of the previous theoretical studies on density fluctuations in time-of-flight imaging rely on the assumption of no atom-atom interactions in the expansion dynamics. This work clearly demonstrated the significance of the interaction effect, in particular for expanding quasi-2D atomic gases. Experimentally, the interaction effect can be suppressed in a deeper 2D regime with tighter axial confinement ($\mu/\hbar\omega_z \ll 1$). Alternatively, one may consider dynamic control of the interaction strength using Feshbach resonances to forcefully turn off the interactions at the beginning of the expansion~\cite{sakharov}. Recently, a quantitative study on the power-law decay of the phase coherence was reported with non-equilibirum exciton-polariton condensates~\cite{Nitsche}.

\begin{acknowledgements} 

We thank Igor Mazets, Tim Langen, and Ludwig Mathey for helpful discussions. This work was supported by the NRF of Korea (Grants No. 2011-0017527, No. 2008-0062257, No. 2013-H1A8A1003984).

\end{acknowledgements}


\begin{references}
\bibitem{footnote1} We observed that thermal samples with optical densities $\leq 0.2$ give the same spectrum of $NP_e(q)$ as in Fig.~3.
\end{references}
\end{document}